\documentclass[a4paper,
               ]{jacow}
\usepackage{pdfpages,multirow,ragged2e} %
\makeatletter%
	\ifboolexpr{bool{xetex}}
	 {\renewcommand{\Gin@extensions}{.pdf,%
	                    .png,.jpg,.bmp,.pict,.tif,.psd,.mac,.sga,.tga,.gif,%
	                    .eps,.ps,%
	                    }}{}
\makeatother

\ifboolexpr{bool{xetex} or bool{luatex}} 
 {}                                      
 {\usepackage[utf8]{inputenc}}           

\usepackage[USenglish]{babel}
\usepackage{hyperref}
\ifboolexpr{bool{jacowbiblatex}}%
 {%
  \addbibresource{jacow-test.bib}
  \addbibresource{biblatex-examples.bib}
 }{}
\begin{document}

\title{Impact of medium temperature heat treatment on flux trapping sensitivity in SRF cavities\thanks{The work is partially supported by the U.S.Department of Energy, Office of Science, Office of High Energy Physics under Awards No. DE-SC 0009960 and  The manuscript has been authored by by Jefferson Science Associates, LLC under U.S. DOE Contract No. DE-AC05-06OR23177.}}
\author{P. Dhakal\textsuperscript{1,2}\thanks{dhakal@jlab.org},   B. D. Khanal\textsuperscript{2}, E. Lechner\textsuperscript{1},  and G. Ciovati\textsuperscript{1,2} \\
				    \textsuperscript{1}Thomas Jefferson National Accelerator Facility, Newport News, VA 23606, USA \\ 
                    \textsuperscript{2}Center for Accelerator Science, Department of Physics, \\Old Dominion University, Norfolk, VA 23529, USA\\
		}
	
\maketitle

\begin{abstract}
The effect of mid-T heat treatment on flux trapping sensitivity was measured on several 1.3 GHz single cell cavities subjected to vacuum annealing at temperature of 150 - 400 $^\circ$C for a duration of 3 hours. The cavity was cooldown with residual magnetic field $\sim$0 and $\sim$20 mG in the Dewar with cooldown condition of full flux trapping. The quality factor as a function of accelerating gradient was measured. The results show the correlation between the treatment temperature, quality factor, and sensitivity to flux trapping. Sensitivity increases with increasing heat treatment temperatures within the range of (200 - 325 $^\circ$C/3h). Moreover, variations in the effective penetration depth of the magnetic field and the density of quasi-particles can occur, influencing alterations in the cavity's electromagnetic response and resonance frequency.
\end{abstract}

\section{Introduction}
The current leading technique in processing superconducting niobium surfaces for SRF cavities involves surface engineering through impurity alloying via thermal treatment. This method, considered state-of-the-art, takes advantage of the introduction of impurities from external sources such as titanium or nitrogen, or the presence of native niobium oxides within the rf penetration depth  \cite{dhakal13, anna, dhakalreview, posen, ito, eric}. However, impurities play a crucial role in increasing the density of pinning centers, enhancing susceptibility to flux trapping during cavity cooldown due to an incomplete Meissner effect. To optimize the process, it is essential to minimize flux trapping sensitivity, resulting in a higher quality factor and an increased accelerator gradient.

A recent advancement in cavity heat treatment, known as "mid-T" baking, capitalizes on the naturally formed oxide on the surface of the Nb cavity \cite{posen,ito, eric, feisi,zhitao}. This oxide is then dissolved in the bulk material. The diffusion of oxygen is temperature and time-dependent, influencing the concentration of interstitial oxygen. Developing processes with lower flux trapping sensitivity is paramount for achieving higher quality factors and accelerator gradients. In this manuscript, we present the results of the several SRF cavity tests subjected to mid-T heat treatment in the furnace. The quality factor as a function of accelerating gradient as well as the flux trapping sensitivity was measured. Additionally, the change in resonant frequency near the critical temperature $T_C$ was measured to confirm the change in electronic properties as a result of heat treatment.

 \section{Experimental setup}
Two 1.3 GHz TESLA shaped single cell cavities, namely TE1-05 and TE1-06 were  used for this current study. The cavities were previously heat treated at higher temperature (1000 $^\circ$C) followed by 25 $\mu$m electropolishing to ensure that the cavity provides good flux expulsion \cite{bashuSRF23}.  Before mid-T heat treatment, the cavities were high pressure rinsed with de-ionized water, dried in a class 10 clean room, installed clean Nb foils over the beamtube flanges. The furnace temperature was increased at a rate of $\sim 2-5~^\circ$ C / min until it reached the target temperature. The furnace was held at the desired temperature for 3 hours and cooled to room temperature. The cavities were subjected to standard cavity processing techniques of high pressure rinse and assembly. Three flux gate magnetometers were taped at the equator 120$^\circ$ apart each and parallel to the cavity axis. The cooldown of cavity was done by maintaining residual magnetic flux density of $\sim$0 and $\sim$20 mG in the Dewar with cooldown condition of full flux trapping. The RF measurements consist of $Q_0(T)$ at low peak rf field $B_p\sim 15~ mT$ and $Q_0(B_p)$ at 2.0 K. Furthermore, the resonant frequency and quality factor were measured during the cavity warm up through $T_c$. After each rf test, the cavity's rf surface is reset by $\sim$25 $\mu$m electropolishing, before next heat treatment.
\section{Experimental Results}

\subsection{rf results}
\begin{figure}[htb]
   \centering
   \includegraphics*[width=0.8\columnwidth]{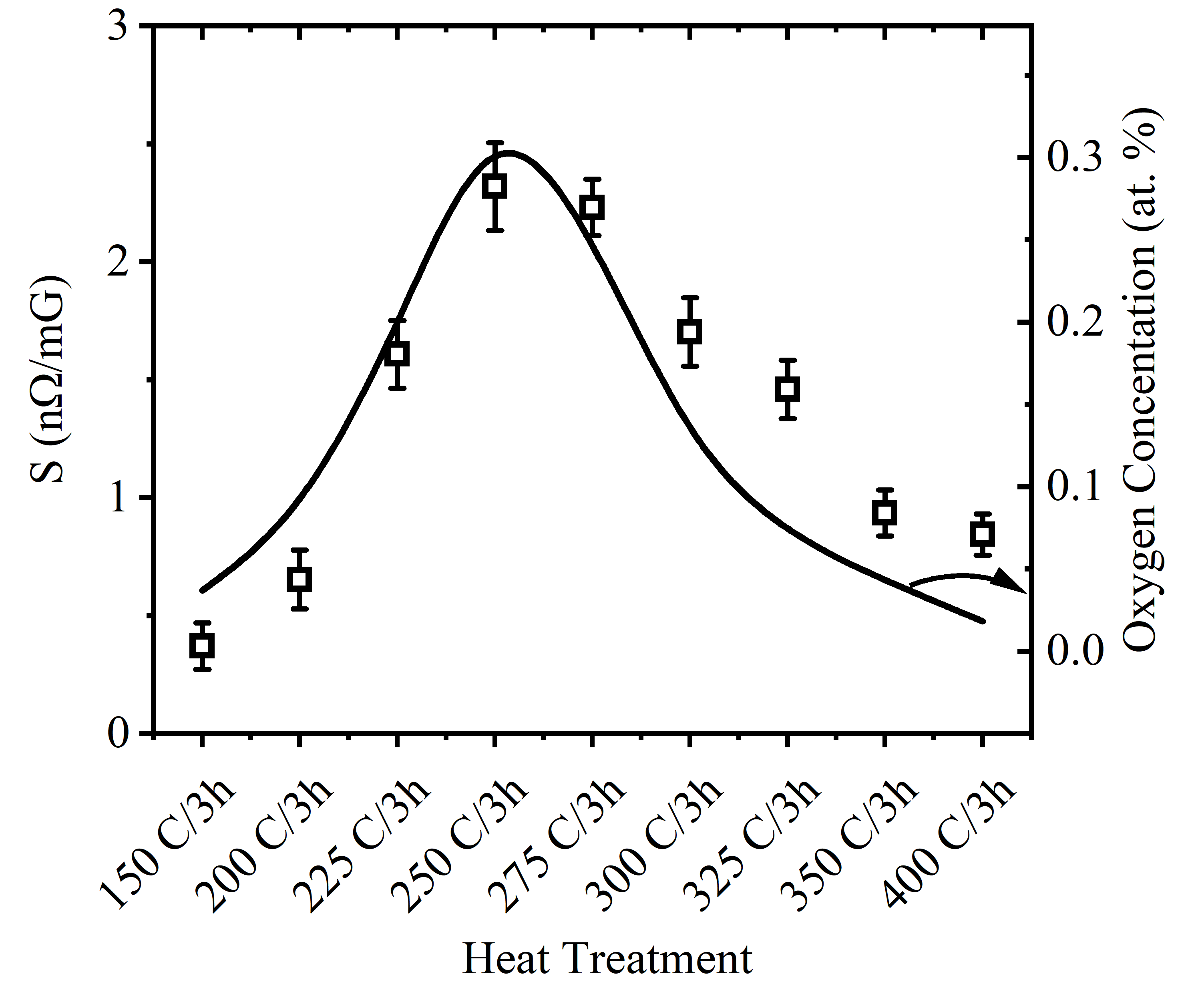}
   \caption{Flux trapping sensitivity as a function of heat treatment temperature. The solid line is calculated oxygen concentration within first 100 nm using oxygen diffusion model \cite{ericjap}.}
   \label{Fig1}
\end{figure}

\begin{figure}[htb]
   \centering
   \includegraphics*[width=0.8\columnwidth]{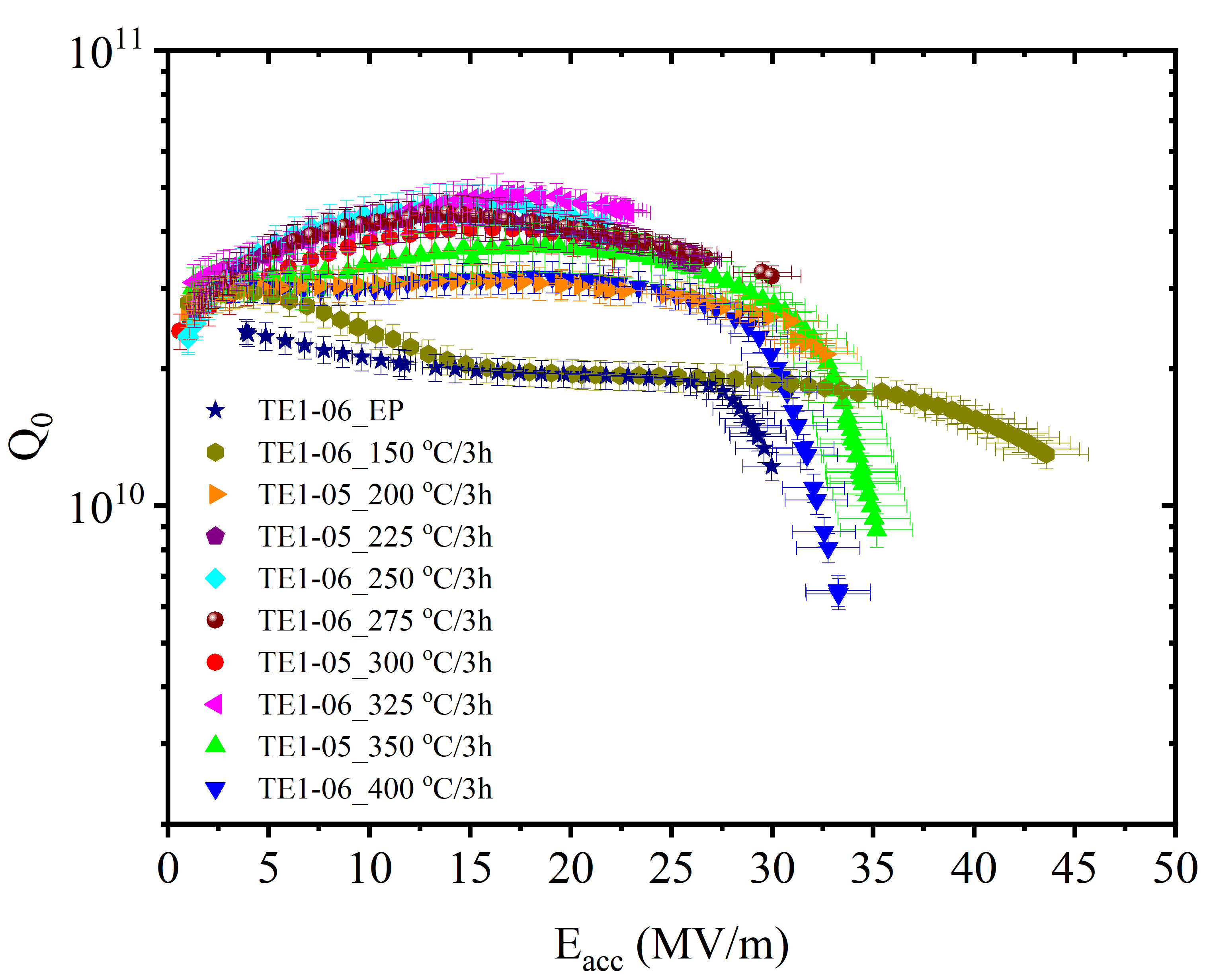}
   \caption{$Q_0$ vs. $E_{acc}$ at 2.0 K with < 2 mG residual field in Dewar.}
   \label{Fig2}
\end{figure}

The $R_s(T)$ data were fitted using the model presented in Ref. \cite{gigi14} to extract the temperature-dependent $R_{BCS}$ and temperature-independent residual resistance $R_i$. The flux trapping sensitivity is calculated using,
\begin{equation}
    S = \frac{R_{i,B_2}-R_{i,B_1}} {B_2-B_1}.
\end{equation}
where $B_1 \approx $ 0 mG and $B_2 \approx $ 20 mG. Figure \ref{Fig1} shows the flux trapping sensitivity as a function of heat treatment temperature. Sensitivity increases with the increase in the temperature of heat treatment with the maximum at $\sim$ 250 $^\circ$ / 3h. On further increase in temperature the flux trapping sensitivity decreases gradually. The flux trapping sensitivity follows the oxygen concentration within 100 nm calculated using oxygen diffusion model \cite{GigiAPL, ericjap}.
Figure \ref{Fig2} shows the $Q_0$ vs. $E_{acc}$ at 2.0 K for both cavities. The baseline cavity was limited to $\sim$ 30 MV/m with a high field Q-slope. In the same cavity, when subjected to 150 $^\circ$C/3 h of heat treatment, the cavity gradient reached to $\sim$ 44 MV/m before it quenched. Eliminating the high-field Q-slope with low-temperature baking has been a proven technique to improve the accelerating gradient \cite{gigijap, bashuieee}. With a further increase in the temperature of the heat treatment, the quality factor increases with increasing accelerating gradient with $Q_0$ reaching $\sim 4 \times 10^{10}$ at $E_{acc} \sim 20$ MV/m for 275 - 325 $^\circ$ C/3h with a lower quench gradient ($\sim 25$ MV / m) compared to the baseline test. On further increase in heat treatment temperature, the overall $Q_0$ gets lower while the high field Q-slope starting at $\sim$ 25 MV/m. 

\begin{figure}[htb]
   \centering
   \includegraphics*[width=0.8\columnwidth]{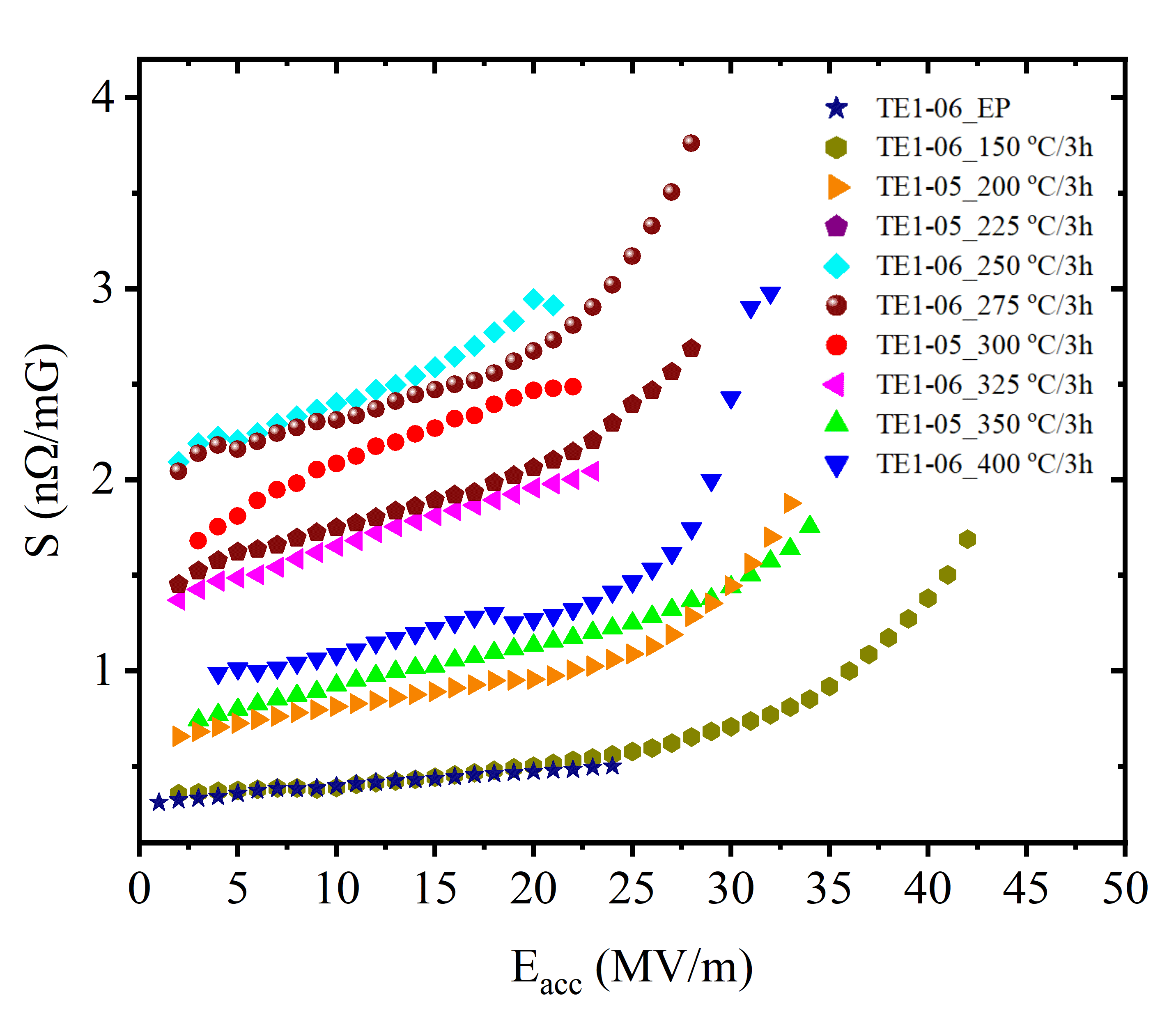}
   \caption{Flux trapping sensitivity as a function of $E_{acc} $ at 2.0 K.}
   \label{Fig3}
\end{figure}

For all tests, the measurements were repeated with $\sim$ 20 mG of residual magnetic flux density trapped during the cavity cooldown. The accelerating gradient dependence of flux trapping sensitivity was calculated using
\begin{equation}
    S = \frac{R_{s,B_2}-R_{s,B_1}} {B_2-B_1}.
\end{equation}
where $R_s = G/Q_0$, $B_1 \approx $ 0 mG and $B_2 \approx $ 20 mG. Figure \ref{Fig3} shows the flux trapping sensitivity as a function of accelerating gradient. As shown in Fig. \ref{Fig1}, the flux trapping sensitivity increase with the increase in heat treatment temperature with highest sensitivity for 250 $^\circ$C/3h heat treatment. On further increase in temperature, the gradient dependence of sensitivity decreases. 

\subsection{$f$ and $Q_0$ during warm up }
\begin{figure}[htb]
   \centering
   \includegraphics*[width=0.8\columnwidth]{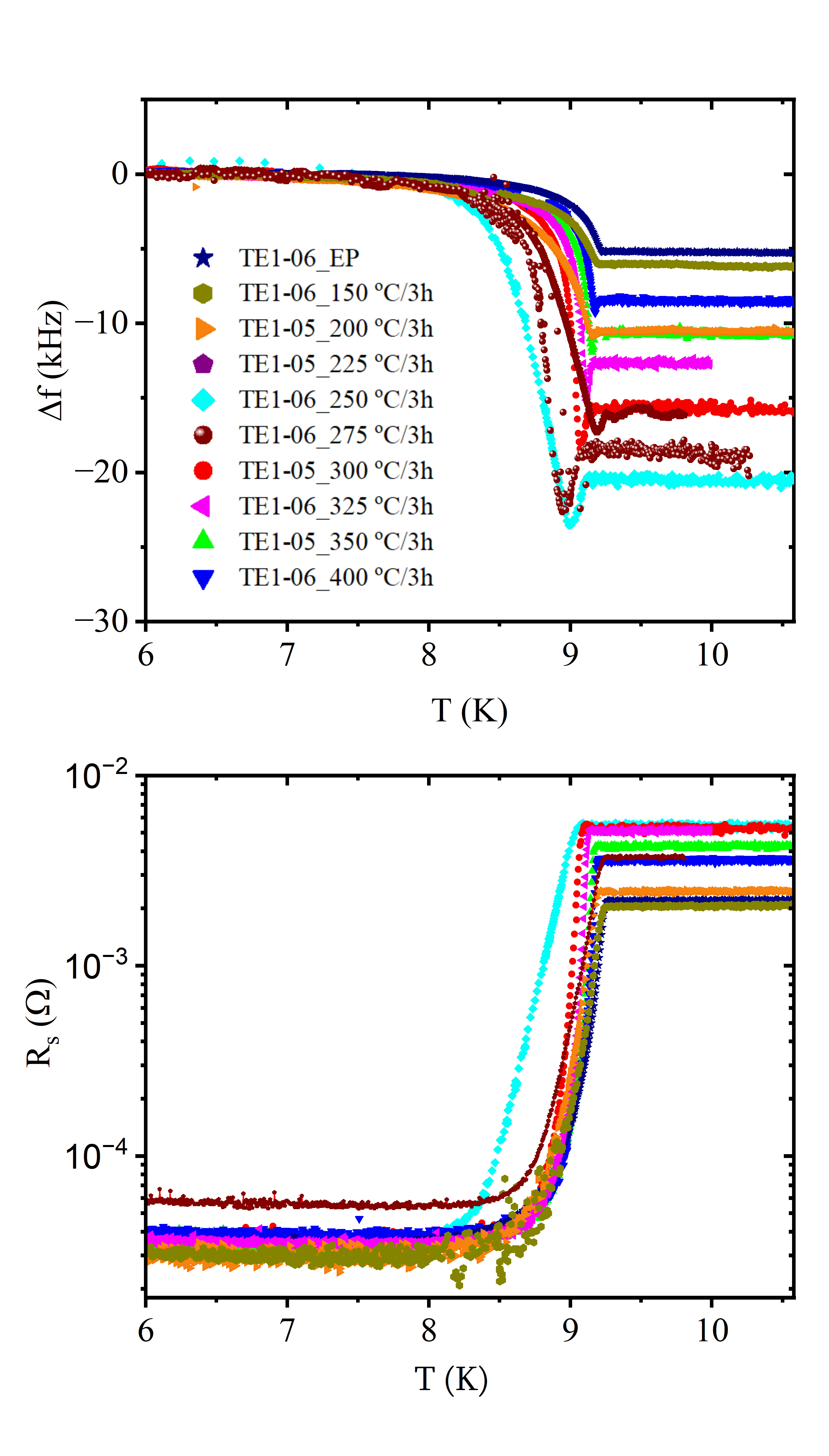}
   \caption{Change in resonant frequency and surface resistance as a function of temperature.}
   \label{Fig4}
\end{figure}

\begin{figure}[htb]
   \centering
   \includegraphics*[width=0.82\columnwidth]{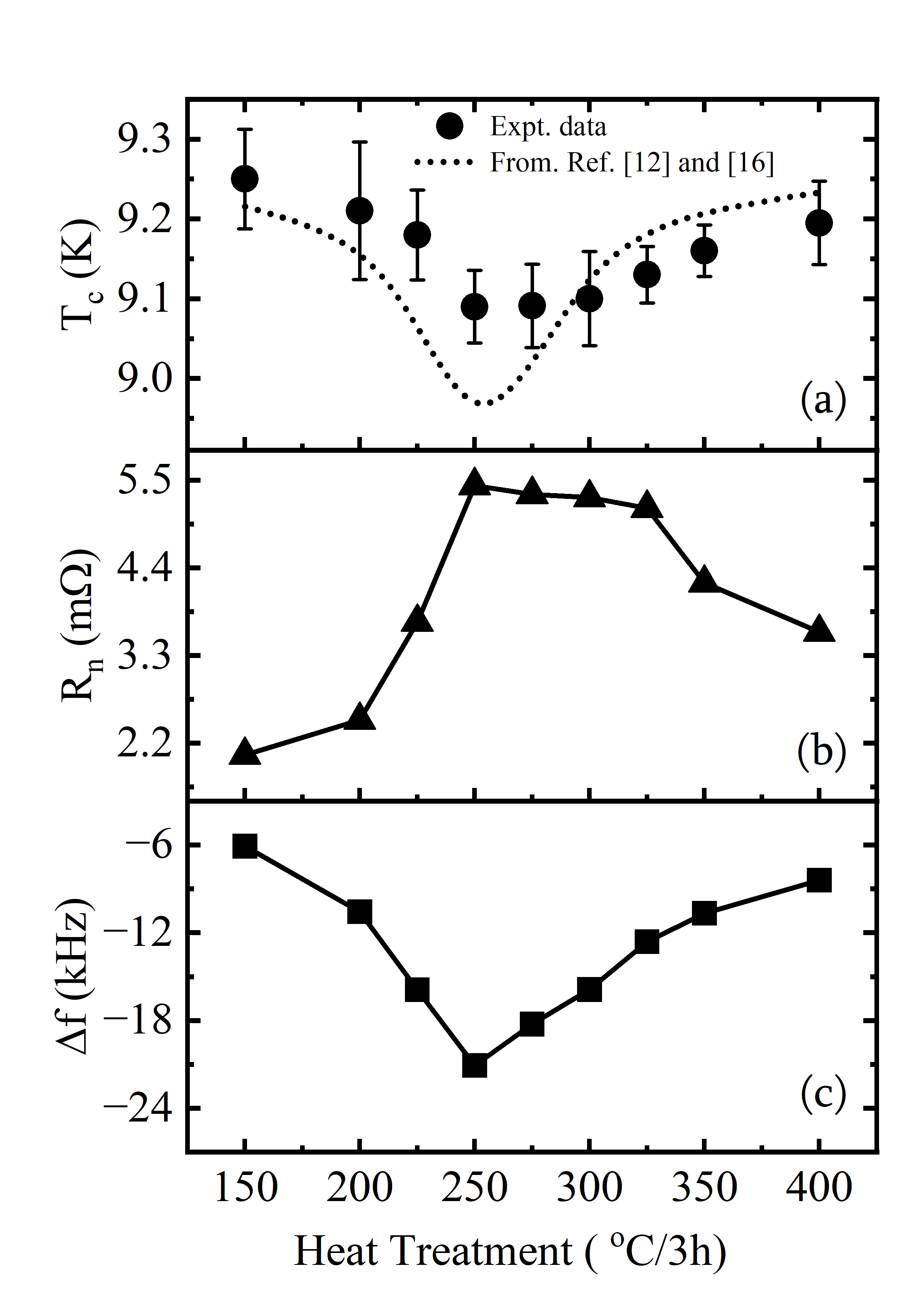}
   \caption{(a) Superconducting transition temperature ($T_c$), (b) surface resistance, $R_n$ at 10 K, and (c) frequency shift ($\Delta f = f_{6~K}-f_{10~K}$) as a function of heat treatment temperature. The dotted line in (a) is calculated  $T_c$ due to solute solution of Nb-O \cite{desorbo} within the first 100 nm using the oxygen diffusion model \cite{ericjap}.}
   \label{Fig5}
\end{figure}

After the completion of the rf test, the cavity was connected to a network analyzer to measure the resonant frequency and quality factor as a function of temperature. To correct the Dewar pressure fluctuation, the pressure sensitivity was corrected by tracking the frequency with respect to the Dewar pressure at a given temperature. All data were corrected for at 1 atm (760 Torr) pressure. Figure \ref{Fig4} shows the change in frequency $\Delta f = f - f_{6 K}$ and $R_s = G/Q_0$ as a function of temperature. The change in $\Delta f$ and $R_s$ is clearly seen as a result of heat treatments. The baseline EP cavity (TE1-05) does not show any dip near $T_c$, whereas both the $\Delta f$ and depth of the dip increase as the heat treatment temperature increases up to $\sim $ 275 $^\circ$C/3h. On further increase in temperature, both $\Delta f$ and depth of the dip decrease, as shown in Fig.\ref{Fig4}. The effect of the current heat treatment process not only changes the frequency behavior, but also has an effect on $R_s$ and $T_c$, as shown in Fig.\ref{Fig5}. 

\section{Discussion and Summary}
The enhancement of the quality factor through Q-rise is evident following heat treatments within the temperature range of 200 to 400 $^\circ$C  over a duration of 3 hours. In particular, a higher $Q_0$ manifests itself when the electronic mean free path is comparable to the superconducting coherence length. The introduction of oxygen reduces the electron mean free path thereby increasing the penetration depth, but decreasing the dissipative conductivity. The results shown here are in qualitative agreement with the a recent characterization of the oxygen diffusion process \cite{eric} based on Ciovati's model\cite{GigiAPL}. It is interesting to note that the baking around 150 $^\circ$C/3hr ameliorated the high field Q slope. In the context of shallow impurity diffusion profiles, where the peak supercurrent density is deformed and pushed away from the surface, this result is in qualitative agreement with the expected optimal baking time and temperature \cite{ericjap}. It should be emphasized that cavities treated within the temperature range of 250 to 325 $^\circ$C/3h exhibit a tendency to quench at lower gradients (20 - 25 MV/m) compared to cavities treated outside of this temperature range. Furthermore, several instances of strong multipacting were observed at this gradient range in contrast to the baseline EP'ed cavities, suggesting a possible surface modification due to heat treatment, which causes a change in secondary electron emission yield.  

As anticipated, heat treatment causes the change in local mean free path as well as spatial distributions of impurities, oxides, dislocation networks \cite{dhakal20}. The sensitivity to flux trapping increases with rising heat treatment temperatures within the range of (200 - 250 $^\circ$C/3h). It is possible that within this temperature range, surface oxide fails to fully dissolve, resulting in an accumulation of higher oxide nanoprecipates near the surface, thus promoting a higher density but weak pinning centers. However, when the temperature exceeds 250 $^\circ$C for the same duration, the breakdown of surface oxides and subsequent diffusion into the bulk decrease the pinning centers on the surface. Consequently, the sensitivity to flux trapping gradually diminishes with increasing heat treatment temperatures, as shown in Figs. \ref{Fig1} and \ref{Fig3}. The observation is further corroborated by the oxygen concentration profile as shown in Fig. \ref{Fig1} with peak oxygen concentration at $\sim$ 250 $^\circ$C/3h heat treatment.

The change in transition temperature as shown in Fig. \ref{Fig5} is qualitatively in agreement with the literature values of reduction in $T_c$ for solute solution of oxygen in niobium \cite{desorbo}. The redistribution of the oxide on the Nb surface due to heat treatment  introduce an additional scattering mechanisms and modify the electronic characteristics of the superconducting material in proximity to the transition temperature. Consequently, variations in the effective penetration depth of the magnetic field and the density of quasi-particles can occur, influencing alterations in the cavity's electromagnetic response and resonance frequency \cite{zarea}. 

\section{ACKNOWLEDGMENTS}
 We would like to acknowledge Jefferson Lab SRF production technical staff members for the cavity fabrication, chemical processing, heat treatment, clean room assembly, and cryogenic support.

%
%
\ifboolexpr{bool{jacowbiblatex}}%
	{\printbibliography}%
	{%
	
	
} 

\end{document}